\documentstyle[preprint,aps,floats,tighten]{revtex}

\input{psfig.tex}
\begin{document}
\draft
\preprint{\vbox{\hbox{CU-TP-806} 
                \hbox{CAL-623}
		\hbox{IASSNS-AST-96/65}
                \hbox{astro-ph/9612209}
}}
\title{The Electron-Screening Correction for the
Proton-Proton Reaction}
\author{John N. Bahcall\footnote{jnb@ias.edu}}
\address{School of Natural Sciences, Institute for Advanced Study, Princeton,
NJ~~08540}
\author{Xuelei Chen\footnote{xuelei@phys.columbia.edu} and
Marc Kamionkowski\footnote{kamion@phys.columbia.edu}}
\address{Department of Physics, Columbia University, 538 West
120th Street, New York, NY~~10027}
\date{December 1996}
\maketitle

\begin{abstract}
We  test the Salpeter formalism for calculating electron
screening of nuclear fusion reactions by solving numerically the
relevant Schrodinger equation for the fundamental proton-proton
reaction.  We evaluate exactly the square of the overlap
integral of the two-proton  wave function and the deuteron wave
function and compare with the usual analytic approximation.  The
usual WKB solution agrees with the numerical solution to
$O(10^{-4})$. 
\end{abstract}

\section{Introduction}

Over the past three decades, much work has been devoted to refining
the input data used in  
calculating  solar-neutrino fluxes.  The comparison between the 
predicted and the observed fluxes has important implications for
particle physics and astrophysics. Most recently, a great deal of
attention has been paid to a relatively minor effect, the 
electron screening of nuclear fusion reactions
\cite{Salpeter,Dzitko,Mitler,Cararo,Langanke,Shaviv}, 
and it has even 
been argued \cite{DarShaviv} that the
discrepancy between observations and theoretical predictions
might be reduced significantly 
if a different  screening correction is adopted. 

We test in this paper the robustness of the standard WKB 
analytic treatment due
to Salpeter \cite{Salpeter} by solving numerically the relevant
Schrodinger equation, including a Debye-Huckel screening potential, 
for the fundamental proton-proton ($pp$) reaction.
The unscreened rate of this reaction can be calculated 
precisely \cite{KamBah}
using standard weak-interaction theory, accurate 
laboratory data for the two-proton system, and
different refined deuteron wave functions in agreement with a variety
of nuclear-physics measurements. Radiative corrections are also 
included in the standard calculation \cite{KamBah}.

In the next Section, we re-derive Salpeter's analytic result for the
weak-screening limit using a kinetic-theory approach (rather
than Salpeter's thermodynamic arguments).  In Section III, we
calculate the proton-proton wave function in both a screened and
unscreened Coulomb potential by numerical solution of the
Schrodinger equation.  We then use these results to evaluate
numerically the electron-screening correction to the
proton-proton reaction, thereby testing the validity of the
standard WKB calculation.  In an Appendix, we calculate a
correction to Salpeter's result and find it to be negligibly
small.

\section{Review of Screening Correction}

To begin, we re-derive Salpeter's screening correction.  To do
so, we use kinetic theory to calculate reaction rates in both a
screened and unscreened plasma.  Although Salpeter's derivation
was based on a thermodynamic argument, this alternative approach
recovers the same results, and it will be useful for
understanding the numerical work in the following Section.
With our analytic approach, a very small correction to Salpeter's
results is 
obtained and presented in the Appendix.

The nuclear fusion rates in the solar interior are controlled primarily by 
Coulomb barriers. Therefore, the energy dependence of the fusion
cross section is usually written as 
\begin{equation}
     \sigma(E)~\equiv~\frac{S(E)}{E}\exp(-2\pi\eta),
\label{phenofit}
\end{equation}
where $S(E)$ is a function that varies smoothly in the absence
of resonances, and
\begin{equation}
\label{eta}
\eta=\frac{Z_1 Z_2 e^2}{\hbar v}.
\end{equation}
Here, $Z_1 e$ and $Z_2 e$ are the charges of the two colliding nuclei, and $v$ 
is their relative velocity.  

The controlling factor, $\exp(-2\pi\eta)$, in
Eq. (\ref{phenofit}) takes into account the probability for
the nuclei to tunnel through the Coulomb barrier.  It is
obtained from the Coulomb potential
$V(r)=Z_1 Z_2 e^2/r$ through the WKB approximation,
\begin{eqnarray}
     \Gamma(E)&=&\exp\left( \frac{-2}{\hbar}
     \int_{0}^{r_c}~~[2\mu(V_{\rm Coul}(r)-E)]^{1/2} dr 
     \right) \nonumber \\
     &=&\exp\left(-2 Z_1 Z_2 \frac{e^2}{\hbar}\sqrt{\frac{2\mu}{E}}\int_{0}^{1}
     \sqrt{\frac{1}{u}-1} ~~du\right)\\
     &=&e^{-2\pi\eta},
\label{gamowfactor}
\end{eqnarray}
where $r_c $ is the classical turning point, defined by $V_{\rm
Coul}(r_c) =E$,
$E$ is the kinetic energy, and $\mu$ is the reduced mass,
\begin{equation}
     \mu=\frac{m_1 m_2 }{m_1 + m_2 }.
\end{equation}
Here, $m_1 $ and $m_2 $ are the masses of the reacting nuclei.

If the energies of the reacting nuclei have a Maxwell-Boltzmann
distribution at a temperature $T$, the thermally-averaged cross
section times relative velocity is \cite{Bahcall}
\begin{equation}
     \langle\sigma v\rangle=\sqrt{\frac{8}{\pi \mu (k_B T)^3}}
     \int_{0}^{\infty} dE ~S(E)
     \exp(-2\pi\eta-E/k_B T).
\label{velocity}
\end{equation}
However, in the stellar interior each nucleus, even though completely ionized, 
attracts neighboring electrons and repels neighboring nuclei; thus, the 
potential between two colliding nuclei is no longer a pure Coulomb potential, 
but a screened potential $V_{\rm sc}(r)$. In the weak-screening case,  
the Coulomb interaction energy between a nucleus and its nearest few 
electrons and nuclei of the gas is small compared with the thermal energy 
$k_B T$.  In this case, the surrounding electrons 
and ions are only slightly displaced, and we obtain a screened
potential of the form \cite{Salpeter}
\begin{equation}
     V_{\rm sc}(r) = \frac{Z_1 Z_2 e^2 }{r}\exp(-r/r_D).
\label{screenedpotential}
\end{equation}
Here, 
\begin{equation}
     r_D= \zeta^{-1/2} \left( {k_B T a \over e^2} \right)^{1/2}
     a ,
\label{debyeradius}
\end{equation}
is the Debye radius for the cloud; 
$A=A_1 A_2 /(A_{1}+A_{2})$ is the reduced mass in atomic mass units;
$\zeta = \sqrt{\sum_{i} (X_i Z_{i}^2 /A_{i} +
X_{i}Z_{i}/A_{i})}$; $X_{i}$, $Z_i$, and $A_i$ are the mass
fraction, charge, and mass number, respectively, of nucleus $i$;
\begin{equation}
     a = \frac{1}{(4\pi\rho N_0 )^{1/3}} = \rho^{-1/3} (0.51 \times
     10 ^{-8}\, {\rm cm})
\end{equation}
is a measure of interparticle distance; $\rho$ is the density in units 
of g~cm$^{-3}$; and $N_0$ is Avogadro's constant. 

For the screened potential, the penetration
factor is then given by 
\begin{equation}
     \Gamma(E)=\exp\left(\frac{-2r_c }{\hbar}\sqrt{2\mu
     E}\int_{0}^{1}\left[\frac{1}{u}
     \exp\left(x(1-u)\right)-1\right]^{1/2}du
\right),
\end{equation}
where $x=x(E)= r_c/r_D$.  Here, $r_c$ is the classical
turning-point radius defined by $V_{\rm sc}(r_c)=E$.  However,
if $x$ is small, then $r_c$ for the screened potential is
roughly that for the unscreened potential: $r_c \simeq Z_1 Z_2
e^2/E$.  By expanding the exponential in the small-$x$ limit (to
be justified below), we obtain
\begin{equation}
\label{ave}
     \Gamma(E)=\exp[-2\pi\eta(1-x/2)]=e^{-2\pi\eta} e^{x\pi\eta}.
\end{equation}
Although $x\pi\eta$ depends on the energy, the effect of 
the correction on the thermally-averaged cross section can be
approximated by evaluating $x\pi\eta$ at the most probable
energy of interaction,
\begin{eqnarray}
\label{Edef}
     E_{0}&=&[(\pi\alpha Z_{1} Z_{2} k_{B}T)^2 (mAc^2 /2)]^{1/3}\\
          &=&1.2204(Z_{1}^{2}Z_{2}^{2}AT_{6}^{2})^{1/3}\; \; {\rm keV}
\end{eqnarray}
where $m$ is the atomic mass unit, and $T_{6}$ is the
temperature in units of $10^6$ K.  Then,
\begin{equation}
     \langle\sigma v\rangle\simeq\sqrt{\frac{8}{\pi \mu (k_B T)^3}}
     f_0 \int_{0}^{\infty} dE S(E)
     \exp(-2\pi\eta-E/k_B T),
\label{integralequation}
\end{equation}
where the Salpeter factor $f_0 $ is given by
\begin{equation}
     f_0 = e^{x_0 \pi\eta}=\exp(0.188Z_1 Z_2 \zeta
     \rho^{1/2}T_{6}^{-3/2}),
\end{equation}
where $\rho$ is the density in units of g~cm$^{-3}$,
and
\begin{equation}
     x_0 = x(E_0) = 0.0133\, (Z_1 Z_2)^{1/3} A^{-1/3} \rho^{1/2}
     T_6^{-7/6} \zeta.
\label{xsubzero}
\end{equation}
For the $pp$ reaction, $x_0\simeq 0.01$, which justifies the
small-$x$ approximation used above.  Equations
(\ref{velocity})--(\ref{xsubzero}) provide an alternative
derivation of the Salpeter \cite{Salpeter} weak-screening formula.

\section{Numerical Results}

We now calculate numerically the cross section for the $pp$
reaction for a Coulomb potential and a screened Coulomb
potential to compare with the WKB calculation of the screening
correction. To do so, we note that the 
reaction rate is proportional to $\Lambda^{2}$ \cite{Bahcalllap}, where
$\Lambda$ is the overlap integral of the proton-proton wave function and 
the deuteron wave function,
\begin{equation}
     \Lambda=\sqrt{\frac{a_{p}^{2}\gamma^{3}}{2}} \int u_{d}(r)
     u_{pp}(r) dr,
\label{Lambdaeqn}
\end{equation}
where $a_{p}$ is $pp$ scattering length, $\gamma=\sqrt{2\mu E_{d}}$ is the
deuteron binding wave number, and $E_d$ is the deuteron binding energy.
The function $u_{d}(r)$ is the radial part of the 
$S$-state deuteron wave function.  Our calculation in this
Section follows the approach and notation of Ref. \cite{KamBah}.

For the purposes of this exercise, we use the McGee wave
function \cite{McGee} for the deuteron.  If another wave function (which fits
 the deuteron data) is used, the overlap integral changes only
slightly.  Since we are here only investigating the effect of
the screening correction to the reaction rate, our specific
choice of the deuteron wave function is unimportant.

The radial wave function $u_{pp}(r)$ satisfies the radial
Schrodinger equation,
\begin{equation}
\label{coulomb}
     \frac{d^{2}u}{dr^2}-\left[\frac{1}{Rr}+V_{\rm
     nuc}(r)\right]u=-k^2 u,
\end{equation}
where
\begin{equation}
   R=\frac{\hbar^2}{2\mu e^2}=28.8198 \; {\rm fm},
\end{equation}
$k=\mu v/\hbar$ is the center-of-mass momentum, and $V_{\rm
nuc}(r)$ is the short-range nuclear potential.  For $V_{\rm
nuc}(r)$ we use an exponential potential which yields the
observed value for the scattering length and effective range
\cite{KamBah}.  Again, the overlap integral turns out to be
practically  independent of the detailed shape of the nuclear
potential (as long as it matches the measured scattering length
and effective range), so the choice of nuclear potential is 
 unimportant for determining the screening
correction.  

In the weak-screening case, Eq. (\ref{coulomb}) is replaced by
\begin{equation}
\label{screen}
     \frac{d^{2}u}{dr^2}-\left[\frac{e^{-r/r_{D}}}{Rr}+V_{\rm
     nuc}(r)\right]u=-k^2 u.
\end{equation}
The solution to the Schrodinger equation is unique once the 
two boundary conditions are given.  The first condition is $u(0)=0$.
The other boundary condition is obtained by noting that the
asymptotic behavior of the wave function for $r \gg r_D$ must be
\cite{Landau},
\begin{equation}
\label{colasy}
     u_{pp}^{\rm coul}(r) \sim N_{\rm coul}\sin \left(kr -
     \frac{1}{2kR} \log(2kr)
     +\delta^{\rm coul} _{0}\right),
\end{equation}
for the Coulomb potential, and 
\begin{equation}
\label{scasy}
     u_{pp}^{\rm sc}(r) \sim N_{\rm sc}\sin(kr -\delta^{\rm sc} _{0}),
\end{equation}
for the screened potential, where $\delta^{\rm coul}$ and
$\delta^{\rm sc}$ are phase shifts.  Fixing the incident flux
of protons for the Coulomb and the screened-Coulomb interactions
requires $N_{\rm coul}=N_{\rm sc}$.

To solve this boundary-value problem, we integrate Eqs.\
(\ref{coulomb}) and (\ref{screen}) from $r=0$ 
with the condition $u(0)=0$ and $u^{\prime}(0)=1$ to a large distance
(about $10\, r_{D}$), and then test that the solutions converge to the form
of Eqs.\ (\ref{colasy}) and (\ref{scasy}), respectively.  From
these numerical solutions, we obtain the normalizations $N_{\rm
sc}$ and $N_{\rm coul}$.  We then use the calculated wave
functions to evaluate the overlap integral in
Eq. (\ref{Lambdaeqn}) both with and without screening.  By
squaring the ratio of the two overlap integrals, we determine
numerically the screening correction to the cross section for
the $pp$ reaction. 

\begin{figure}[htbp]
\centerline{\psfig{file=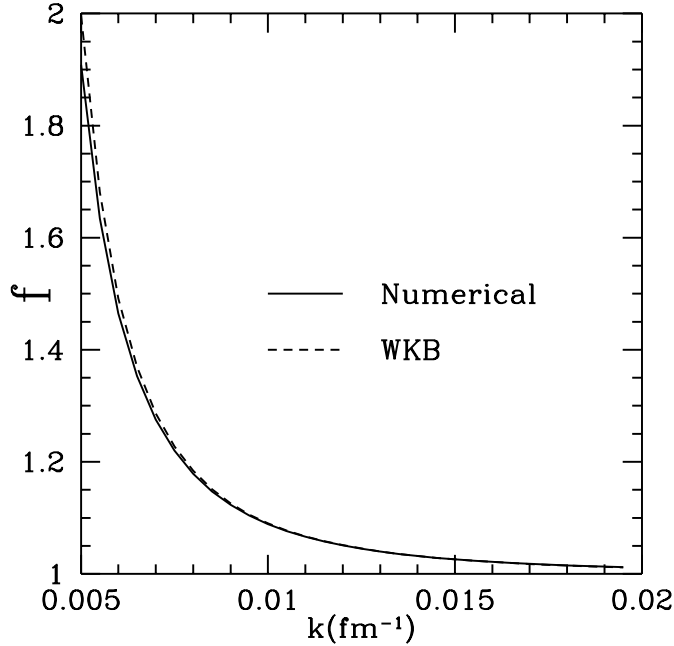,width=3.5in,height=3.5in}}
\bigskip
\caption{\noindent The screening enhancement factor as a
function of $k$.  The dashed curve is that for the WKB result, and
the solid curve is that for the numerical result.
}
\label{numplot}
\end{figure}

In Fig.~\ref{numplot} we plot the WKB (dashed curve) and
numerical (solid curve) results for the screening correction
for the $pp$ reaction as a function of relative momentum $k$.  Our
results show that the discrepancy is small.  For values of $k$
at which the $pp$ reaction occurs ($k \simeq 0.016 \,{\rm
fm}^{-1}$), the fractional difference is $O(10^{-4})$.  In fact,
at smaller $k$, the fractional difference is expected to be even
smaller, as argued below.  The increased discrepancy at smaller
$k$ shown in Fig.~\ref{numplot} is due to numerical error in our
calculation: accurate integration of the Schrodinger equation
becomes increasingly difficult at smaller $k$ since the
asymptotic forms in Eqs.\ (\ref{colasy}) and (\ref{scasy}) are
reached at progressively larger $r$.

\section{Conclusion}

Our main result is that a numerical solution of the screened
Schrodinger equation for the proton-proton reaction reaction
gives results in excellent agreement [to $O(10^{-4})$] with the
rate calculated analytically using the usual WKB approximation,
as originally formulated by Salpeter.

\acknowledgments

This work was supported by D.O.E. contract DEFG02-92-ER 40699,
NASA NAG5-3091, and the Alfred P. Sloan Foundation at Columbia.
JNB was supported at the Institute for Advanced Study by NSF grant
PHY95-13835. 
MK thanks J. Applegate for useful conversations and JNB thanks 
A. Gruzinov for valuable discussions. This work was initiated in
response to a conjecture by A. Dar that the standard treatment of
screening could cause an error of order 5\% in the proton-proton
reaction rate in the sun. We are grateful to A. Dar for this
stimulating discussion. 

\vskip 1cm
\appendix\centerline{\small\bf APPENDIX: A SMALL CORRECTION}
\medskip

In this Appendix, we calculate a correction to Salpeter's
screening formula and find it to be negligibly small.  The
integral in Eq. (\ref{integralequation}) is usually evaluated by
expanding in a power series of the inverse of a large quantity $\tau$,
\begin{equation}
     \tau = 3 E_0 /k_{B}T = 42.487(Z_{1}^{2}Z_{2}^{2}AT_{6}^{-1})^{1/3}.
\end{equation}
The average product can then be written 
in a compact form \cite{Bahcall}:
\begin{equation}
     \langle\sigma v\rangle = 1.3005 \times 10^{-15} \left[
     \frac{Z_{1}Z_{2}}{AT_{6}^{2}}\right]
     ^{1/3} f_0 S_{\rm eff} \exp(-\tau) \; \; {\rm cm}^3 {\rm s}^{-1}
\label{standardthermal}
\end{equation}
where $S_{\rm eff}$ is expressed in keV-barns. To first order in
$\tau^{-1}$ \cite{Bahcall66},
\begin{equation}
\label{Seff0}
     S_{\rm eff}=S(E_{0})\left(1+\tau^{-1}\left[\frac{5}{12}+\frac{5
     S^{\prime}E_{0}}
     {2S}+\frac{S^{\prime\prime}E_{0}^{2}}{S}\right]_{E=E_{0}}\right).
\end{equation}
Expressing the various quantities in terms of their values at $E=0$,
we find \cite{Bahcall66} 
\begin{equation}
     S_{\rm eff}(E_0) \simeq S(0)\left[
     1+\frac{5}{12\tau}+\frac{S^{\prime}(E_{0}+ 
     \frac{35}{36}k_{B}T)}{S}+\frac{S^{\prime\prime}
     E_{0}}{S}\left(\frac{E_{0}}{2}
     +\frac{89}{72}k_{B}T\right)\right]_{E=0}.
\end{equation}

More accurately, however, we should include the factor $e^{x\pi\eta}$ in
the thermal-average integral, Eq. (\ref{ave}). To do so, we rewrite the 
integral as
\begin{equation}
     \langle\sigma v\rangle=\sqrt{\frac{8}{\pi \mu (k_B T)^3}}
     \int_{0}^{\infty} dE S(E)
     \exp(-2\pi\eta-E/k_{B} T + x\pi\eta).
\end{equation}
Introducing the dimensionless quantity $z=E/E_{0}$, the
exponential can be written as
\begin{equation}
     -2\pi\eta-E/k_{B}T+x\pi\eta=-\frac{2\tau}{3}z^{-1/2}-\frac{\tau}{3}z+
     \frac{x_0\tau}{3}z^{-3/2}.
\end{equation}
To first order in $x_0$, the minimum point of the exponent is thus shifted 
to $z=1-x_0$, or $E=E_{0}(1-x_0)$. Using Laplace's method \cite{Bender}
for asymptotic expansion of integrals, we find that the only
$O(x_0)$ correction to Eq.~(\ref{standardthermal}) is in the
expression for $S_{\rm eff}$.  To order $O(\tau^{-1}, x_0)$, it is
\begin{equation}
     S_{\rm eff}=S(E)\left(1+\tau^{-1}\left[\frac{5}{12}
     +\frac{5 S^{\prime}(E)E_{0}}
     {2S(E)}+\frac{S^{\prime\prime}(E)
     E_{0}^{2}}{S(E)}\right]_{E=E_{0} (1-x)}\right). 
\end{equation}
In other words, to first order in $x_0$, $E_{0}$ in
Eq. (\ref{Seff0}) should be replaced by $E_{0}(1-x_0)$. 
Expressed as $S(0)$, we have
\begin{equation}
     S_{\rm eff} \simeq S(0)\left[ 1+\frac{5}{12\tau}
     +\frac{S^{\prime}(0)(E_{0}(1-x)+
     \frac{35}{36}k_{B}T)}{S(0)}+\frac{S^{\prime\prime}(0)E_{0}}{S(0)}
     \left(\frac{E_{0}}{2}(1-2x)
     +\frac{89}{72}k_{B}T\right)\right],
\end{equation}
where we have neglected terms of order $O(x_0/\tau)$.
We see that there is an 
$O(x)$ correction to the $S^{\prime}$ and $S^{\prime\prime}$ terms.
Since $x$ is small ($\sim 10^{-2}$ for the $pp$ reaction at the core of 
the Sun), and the $S'$ and $S''$ terms are generally small
compared with the lowest-order term, these corrections are very
small, typically $\sim 0.1$\%.
Therefore, the standard multiplicative correction factor
($f_0$) should give a screened interaction rate
which is accurate to $O(1\%)$ in the weak-screening
regime.  Furthermore, since $x_0$ increases only very slowly with
increasing mass number, the standard correction should also be
 accurate for other fusion reactions which are in the
weak-screening regime.

\end{document}